\def\fun#1#2{\lower3.6pt\vbox{\baselineskip0pt\lineskip.9pt
  \ialign{$\mathsurround=0pt#1\hfil##\hfil$\crcr#2\crcr\sim\crcr}}}
\relax
\documentstyle[12pt]{article}
\advance\textheight by \topskip
\begin{document}
\vspace{-0.2cm}
\begin{flushright}
SU-ITP-94-12\\
UMHEP-407\\
QMW-PH-94-12\\
hep-th/9406059
\end{flushright}
\vspace{-0.2cm}
\begin{center}
{\large\bf SUPERSYMMETRY AND STATIONARY SOLUTIONS\\
\vskip 0.8 cm
IN DILATON-AXION GRAVITY}\\
\vskip 1.2 cm
{\bf Renata Kallosh${}^{a}$\footnote{E-mail address:
{\tt kallosh@physics.stanford.edu}},
 David Kastor${}^{b}$\footnote{E-mail address:
{\tt kastor@phast.umass.edu}},
Tom\'as Ort\'{\i}n${}^{c}$\footnote{
E-mail address: {\tt ortin@qmchep.cern.ch}}
and Tibor Torma${}^{b}$\footnote{E-mail address: {\tt
kakukk@phast.umass.edu}}
}
\vskip 0.05cm
${}^{a}$Physics Department, Stanford University, Stanford CA 94305,
USA\\
\vskip 0.4 cm
${}^{b}$Department of Physics and Astronomy, University of
Massachusetts,\\
Amherst
MA  01003\\
\vskip 0.4cm
${}^{c}$Department of
Physics, Queen Mary and Westfield College, Mile End  \\
Road, London E1
4NS, U.K.\\
\vskip 0.7 cm
\end{center}
%
\vskip 0.6 cm

\centerline{\bf ABSTRACT}

\begin{quotation}
\vskip -.5 cm
New stationary solutions of $4$-dimensional dilaton-axion gravity are
presented, which correspond to the charged Taub-NUT and
Israel-Wilson-Perj\'es (IWP) solutions of Einstein-Maxwell theory.
The charged  axion-dilaton Taub-NUT solutions are shown to have a number of
interesting
properties:\,   i) manifest $SL(2,R)$ symmetry,\,  ii) an infinite throat in
an extremal limit, \,  iii) the throat limit  coincides with an
exact CFT construction.

The IWP solutions are shown to admit supersymmetric Killing
spinors, when embedded in $d=4, \, N=4$  supergravity. This poses a problem for
the interpretation of supersymmetric rotating solutions as physical ground
states.  In the context of
$10$-dimensional geometry, we show that dimensionally lifted versions of the
IWP
solutions are dual to certain gravitational  waves  in  string theory.

\end{quotation}


\newpage


\section{Introduction}

A good deal is now known about static black hole solutions in the
dilaton-axion theory of gravity, which arises in the low energy limit of
string theory.  Solutions have been found \cite{G}, \cite{STW}, \cite{KOSTATIC}
which
correspond to the Reissner-Nordstr\"om black holes of ordinary
Einstein-Maxwell theory and also to the Majumdar-Papapetrou (MP)
multi-black hole solutions.  Analogues of the charged C-metrics, which
describe pairs of black holes accelerating away from one another, have
been found as well \cite{CMETRIC}.  It is natural to expect that this
correspondence between solutions should continue to hold in the more
general stationary case.  Here, the only solutions known thus far are the
basic rotating black hole solutions with either pure electric or pure
magnetic charge \cite{SEN}.  In this paper we present two additional
classes of stationary solutions to dilaton-axion gravity, which may be
considered to be of black hole {\it type}\footnote{We will use the
annotation black hole {\it type} to encompass natural families of
solutions, which include, in addition to black holes, such things as
naked singularities and spacetimes carrying NUT charge.} . These new
solutions correspond to the charged Taub-NUT solutions \cite{BRILL} and
the Israel-Wilson-Perj\'es (IWP) solutions \cite{IW} of ordinary
Einstein-Maxwell theory.

We find a compact form for the general charged Taub-NUT solution, which
reduces to the form of the charged black hole solution given in
\cite{KOSTATIC}, when the NUT parameter is set to zero.  The charged
Taub-NUT solution in Einstein-Maxwell theory has no curvature
singularities.  In contrast, the dilaton-axion solution is singular, if
the charge exceeds a certain critical value.  In the limit of zero NUT
parameter, this critical charge vanishes and the singularity corresponds
to the singular inner horizon of the charged black holes.  There is thus
a competition between the NUT parameter and the charge in determining
the global structure of these solutions.  The charged Taub-NUT solution
in Einstein-Maxwell theory has an extremal limit in which an infinite
throat arises.  We find that a similar infinite throat exists for the
rescaled string metric of certain extremal Taub-NUT solutions in dilaton-axion
gravity, which have predominantly magnetic charge.  Again, when the NUT
parameter is set to zero, this reproduces a known property of extremal
magnetically charged black holes in this theory \cite{G}.  This throat
solution, with nonzero NUT parameter, was found recently in an exact
conformal field theory construction by Johnson \cite{JOHNSON}.

The original IWP solutions describe collections of charged objects in a
state of equipoise, the charges being such that each object satisfies a
`no force' condition with respect to all of the others.  Amongst the IWP
solutions are the static MP solutions, which describe collections of
extremally charged black holes.  The simplest example of a nonstatic IWP
solution is the Kerr-Newman solution with arbitrary angular momentum per
unit mass $a$, charge $Q$, and mass $M$ satisfying
\begin{equation}\label{extremal}
Q^2 =M^2\, .
\end{equation}
For $a=0$, this is the same as the extremality condition, the maximum
charge that a static black hole can have.  For $a\ne 0$, however, the
extremality condition is
\begin{equation}\label{extremal2}
Q^2  =M^2-a^2\, ,
\end{equation}
so the Kerr-Newman solution in the IWP limit ({\it i.e.} satisfying
(\ref{extremal})), with $a\ne0$, actually describes a naked singularity, rather
than a black hole.  Another simple example of a non-static IWP
solution is the extremal charged Taub-NUT solution mentioned above.  In
this case, as with all Taub-NUT metrics, the inverse metric is singular
unless the time direction is taken to be periodic.  These then also fail
to be black holes.  More generally, Hartle and Hawking \cite{HH} have
shown that amongst the IWP solutions, only the MP solutions describe
black holes.  All of the more general stationary solutions have naked
singularities or other pathological features.

This result is troubling because it has been shown that the IWP
solutions admit Killing spinors of $N=2$ supergravity \cite{T} and,
hence, should be considered supersymmetric ground states of the theory.
The general IWP solutions, however,
do not seem like ground states.  For rotating black holes, the extremal
ones satisfying (\ref{extremal2}) and having vanishing Hawking
temperature, would seem altogether better suited for the title of
physical ground state.  It was recently shown \cite{TRACE} that the
nonstatic IWP solutions fail to be supersymmetric at the quantum level,
due to the trace anomaly of $N=2$ supergravity.  Thus, they are, in this
sense, `less supersymmetric' than the static solutions, failing to be
ground states once quantum corrections are taken into account.  This
provides a possible resolution to the paradox of having unphysical
supersymmetric ground states in the Einstein-Maxwell system.

The IWP solutions of dilaton-axion gravity, which we present below,
share many of the properties of the ordinary IWP solutions, including
their various pathologies.  Dilaton-axion gravity, however, may be
embedded in $N=4$ supergravity and has vanishing trace anomaly.  We show
that the new IWP solutions are again supersymmetric ground states,
admitting Killing spinors of $N=4$ supergravity.  Hence, the possible
resolution of ref. \cite{TRACE} fails in this case, and, if one still
would like to associate the unbroken supersymmetry of asymptotically
flat spacetimes with the absence of naked singularities, one is faced
with a new challenge.

A completely different interpretation of the new IWP solutions arises in
the context of embedding into the 10-dimensional geometry of critical
superstring theory.  It was recently shown \cite{BKO2} that
4-dimensional extremal black holes embedded in a certain way in
$10$-dimensional geometry are dual to a set of supersymmetric string
wave (SSW) solutions in $d=10$, $N=1$ supergravity \cite{BKO}.  Here,
we show that a similar construction can be used to relate the stringy
IWP solutions to a class of SSW's.  In addition to showing that these
solutions are identical from the point of view of string propagation,
this gives an independent demonstration of the supersymmetry of the IWP
solutions.


\section{Taub-Nut Solutions in Dilaton-Axion Gravity}

We follow the conventions of references \cite{US}, \cite{KOSTATIC}.
The signature of the metric is $(+---)$.  The action is given by
\begin{equation}\label{action}
S  =
\frac{1}{16\pi}
\int d^{4}x\sqrt{-g}\biggl \{-{\cal R}+2(\partial\phi)^{2}
+\frac{1}{2}e^{4\phi}(\partial a)^{2}
-e^{-2\phi}F^{2}
+iaF{}^{\star}F\biggr \}\; ,
\end{equation}
where $a$ is the axion, $\phi$ is the dilaton, ${\cal R}$ is the scalar
curvature and $F$ is the field strength\footnote{The spacetime duals are
${}^{\star}F^{\mu\nu}= \frac{1}{2\sqrt{-g}} \epsilon^{\mu\nu\rho\sigma}
F_{\rho\sigma}$, with flat $\epsilon^{0123}= -\epsilon_{0123}= +i$ and
curved $\epsilon^{\hat{0}\hat{1}\hat{2}\hat{3}}= \frac{1}{g}
\epsilon_{\hat{0}\hat{1}\hat{2}\hat{3}}=+i$.  We will sometimes use a
flat 3-dimensional $\epsilon_{ijk}$ such that $\epsilon_{123}=+1$.} of a
$U(1)$ gauge field $A_\mu$.  It is useful to define the complex scalar
field $\lambda=a+ie^{-2\phi}$ and the $SL(2,R)$-dual to the field
strength
\begin{equation}
\tilde{F}=e^{-2\phi}\,{}^{\star}F -iaF \; .
\end{equation}
In terms of these fields, the action reads
\begin{equation}\label{action2}
S=
\frac{1}{16\pi}
\int d^{4}x\sqrt{-g}\biggl
\{-{\cal R}+\frac{1}{2}\frac{\partial_{\mu}\lambda
\partial^{\mu}\overline{\lambda}}{({\mbox{Im}} \; \lambda)^{2}}
-F_{\mu\nu}{}^{\star}
\tilde{F}^{\mu\nu} \biggr \}\; .
\end{equation}
The advantage of using $\tilde{F}$ is that the equations of motion imply
the local existence of a vector potential $\tilde{A}$, satisfying
\begin{equation}
\tilde{F}=i\, d\tilde{A}\; .
\end{equation}
If $A_{t}$ plays the role of the electrostatic potential, then
$\tilde{A}_{t}$ plays the role of magnetostatic potential.

The action (\ref{action}) describes the bosonic part of the
10-dimensional effective action of string theory, dimensionally reduced
to four dimensions.  The 4-dimensional vector field in this theory may
be either an Abelian part of a Yang-Mills multiplet or it may come
from the non-diagonal component of the metric in extra dimensions,
as in Kaluza-Klein theory, and/or
from components of the 2-form gauge field.
Only when the supersymmetric embedding of the
action (\ref{action}) has been specified, can one attribute a definite
origin of the vector field.  In particular, in sec. 4 we will consider
the embedding of the action (\ref{action}) into $N=4$ supergravity
without additional matter multiplets.  In terms of the 10-dimensional
theory, this means that the 4-dimensional vector field arises from
non-diagonal components of the metric, $g_{\mu 4}$, and also from the
corresponding components of the 2-form gauge field $B_{4\mu}=
-g_{4\mu}$, where $\mu = 0,1,2,3$ and $4$ stands for one of the
six compactified  directions.

The metric of the Taub-NUT solution with charge $Q$ in Einstein-Maxwell
theory \cite{BRILL} is given by
\begin{eqnarray}\label{uncharged}
ds^2 &=& f(r)(dt + 2l\cos\theta d\varphi)^2 -f^{-1}(r)dr^2
-(r^2+l^2)d\Omega^2 \, ,\\
f(r) &=& \frac{r^2-2mr-l^2 +Q^2}{r^2 +l^2} \, .\nonumber
\end{eqnarray}
The Taub-NUT spacetimes have no curvature singularities.  The metric,
however, does have so-called `wire' singularities along the axes
$\theta=0$, $\theta=\pi$ on which the metric fails to be invertible.
Misner \cite{MISNER} has shown that the wire singularities may be
removed by making the time coordinate periodic.  This may be seen in the
following way.  To make the metric regular at the north ($\theta=0$) and
south ($\theta=\pi$) poles, we take separate coordinate patches.  Define
new time coordinates $t_N$, $t_S$ on the patches by
\begin{equation}
t=t_N - 2l\varphi = t_S + 2l\varphi \, .
\end{equation}

If $\varphi$ is to be an angular coordinate with period $2\pi$, then for
consistency on the overlaps we must take the time coordinates to be
periodic, with period $8\pi l$.  The surfaces of constant $r$ then turn
out to have $S^3$ topology.

The metric function $f(r)$ in (\ref{uncharged}) has roots at $r_\pm =
m\pm\sqrt{m^2+l^2 -Q^2}$.  For $r>r_{+}$ and $r<r_{-}$ the metric has
closed time-like curves.  Thus, although the form of the the metric is
similar to Schwarzschild, no black hole interpretation is possible.  For
$r$ in the range $r_-<r<r_+$, the coordinate $t$ is spacelike and $r$ is
timelike.  This region describes a nonsingular, anisotropic, closed
cosmological model.  It can be thought of
as a closed universe containing electromagnetic and gravitational
radiation having the longest possible wavelength \cite{BRILL}.  There is
a limit of (\ref{uncharged}), with $Q^2=m^2+l^2$, in which the two roots
$r_\pm$ coincide at $r=m$.  We will call this the extremal limit.  If we
let $x=r-m$ in this limit, then near $x=0$ the metric becomes
\begin{equation}
ds^2\approx \frac{x^2}{m^2 +l^2}\left(dt +2l\cos\theta
d\varphi\right)^2 - \frac{m^2+l^2}{x^2}dx^2 - (m^2 +l^2)d\Omega^2\, ,
\end{equation}
which has the form of an infinite spatial throat of constant
cross-sectional area, though it does not have the simple direct product
form of the $l=0$ throats.

The uncharged Taub-NUT solution, eqn. (\ref{uncharged}) with $Q=0$,
continues to be a solution in dilaton-axion gravity.  The charged
solution could in principle be found from the uncharged one using
solution generating techniques (see {\it e.g.} refs. \cite{CHARGE} ),
but this would most likely lead to a relatively unwieldy form of the
solution.  Rather, by using less direct methods, we have found the
general charged Taub-NUT solutions in dilaton-axion gravity in a form
which naturally extends that given for the general charged black hole
solutions in ref. \cite{KOSTATIC}.  This new solution is given by
\begin{eqnarray}\label{taubnut}
ds^2 &=& f\left( dt + 2l\cos\theta d\varphi\right)^2 - f^{-1} dr^2 -
R^2 d\Omega^2 \, ,
\nonumber \\
\nonumber \\
f & = & \frac{(r-r_+)(r-r_-)}{R^2},\qquad R^2=r^2+l^2 -|\Upsilon|^2\, ,
\nonumber \\
\nonumber \\
r_\pm &=& m\pm r_0,\qquad r_0^2=m^2 +l^2 +|\Upsilon|^2 -4|\Gamma|^2 \, ,
\\
\nonumber\\
\lambda & = & \frac{\lambda_0(r+il) +\bar{\lambda}_0 \Upsilon} {(r+il)
+\Upsilon},
\qquad
\Upsilon = -2 \frac{\bar{\Gamma}^2}{M}\, ,
\qquad
M=m+il\, ,
\nonumber\\
\nonumber \\
A_t & = & \pm\frac{e^{\phi_0}}{R^2}\left\{ \Gamma (r+il+\Upsilon)
+c.c.\right\} \, ,
\nonumber\\
\nonumber \\
\tilde{A}_t & = & \mp\frac{e^{\phi_0}}{R^2}\left\{ i\Gamma
[\lambda_0(r+il)+\bar{\lambda_0}\Upsilon]
+c.c.\right\}\, .
\nonumber
\end{eqnarray}
The solution depends through the two complex parameters
$\Gamma=(Q+iP)/2$ and $M=m+il$ on four real parameters; electric charge
$Q$, magnetic charge $P$, mass $m$ and NUT charge $l$.

The equations of motion of dilaton-axion gravity have an $SL(2,R)$-symmetry,
which was used in \cite{STW} to generate black holes with combined electric
and magnetic charge, and hence nontrivial axion field, from those with either
pure electric or magnetic charge.  In ref. \cite{KOSTATIC} the generally
charged
black hole solution was presented in a compact form which had the
property of manifest $SL(2,R)$ symmetry.
The dilaton-axion charged Taub-Nut solution (\ref{taubnut}) is already in such
a manifestly symmetric form.
In the present context, this means the following. The
$SL(2,R)$ symmetry of the equations of motion of dilaton-axion gravity
allows one, in
principle, to generate new solutions from a known one by applying an
$SL(2,R)$-rotation.  However, our solutions already describe the
whole  $SL(2,R)$-family of solutions. It suffices to perform an
$SL(2,R)$ rotation only on the parameters describing the solution. In
particular, one has to  substitute the $SL(2,R)$-rotated values of the
axion-dilaton field at infinity as well as the $SL(2,R)$-rotated values
of the charges
\begin{equation}
\lambda_{0}^{\prime}  =
\frac{\alpha\lambda_{0}+\beta}{\gamma
\lambda_{0}+\delta}\; ,
\hspace{2cm}
\Upsilon^{\prime}   =
e^{-2iArg(\gamma\lambda_{0}+\delta)}\Upsilon\; ,
\end{equation}
\begin{equation}
\Gamma^{\prime}=e^{+iArg(\gamma\lambda_{0}+\delta)}
\Gamma\; . \label{eq:transhair}
\end{equation}
where $\alpha, \beta, \gamma$ and $\delta$
are the elements of a $SL(2,R)$ matrix
\begin{equation}
R=
\left(
\begin{array}{cc}
\alpha & \beta  \\
\gamma & \delta \\
\end{array}
\right)\; .
\label{SL}\end{equation}
This gives the $SL(2,R)$-rotated solution.

The solutions (\ref{taubnut}) reduce to known solutions in a number of
limits.  If the complex electromagnetic charge $\Gamma$ is set to zero,
then the dilaton-axion charge $\Upsilon$ vanishes as well, and
(\ref{taubnut}) reduces to the uncharged Taub-NUT solution.  When the
NUT charge $l$ is taken to zero, (\ref{taubnut}) reduces to the general
static charged black hole solution of ref. \cite{KOSTATIC}.

There is an extremal limit of (\ref{taubnut}) in which $r_+=r_-=m$.
This corresponds to fixing the parameters such that $2|\Gamma|^2=|M|^2$,
which in turn implies that $|\Upsilon|^2=|M|^2$.  If we write the metric
in terms of a shifted radial coordinate $\bar{r}=r-m$, then in the
extremal limit it becomes
\begin{equation}\label{extremaltn}
ds^2 = \left(1+\frac{2m}{\bar{r}}\right)^{-1}\left(dt +2l\cos\theta
d\varphi\right)^2
-\left(1+\frac{2m}{\bar{r}}\right)\left (d\bar{r}^2 +\bar{r}^2d
\Omega^2\right)\, .
\end{equation}
We will see below that this is an example of a dilaton-axion IWP
metric.

The metric (\ref{extremaltn}) does not have an infinite throat near
$\bar{r}=0$.  However, an infinite throat does arise in the rescaled
string metric $d\bar{s}^2=e^{2\phi}ds^2$ for certain cases.  To see
this, fix the asymptotic value of the dilaton-axion field to
$\lambda_0=i$, for simplicity.  The dilaton field in (\ref{taubnut})
then becomes
\begin{equation}\label{dilaton}
e^{-2\phi} = \frac{R^2}{|r+il+\Upsilon|^2}\  .
\end{equation}
The string metric corresponding to (\ref{taubnut}) is then given by
\begin{equation}\label{stringmetric}
d\bar{s}^2 = |r+il+\Upsilon|^2 \left\{
\frac{(r-r_+)(r-r_-)}{R^4}(dt + 2l\cos\theta d\varphi)^2
-\frac{1}{(r-r_+)(r-r_-)}dr^2 -d\Omega^2\right\} \ .
\end{equation}

Now take the extremal limit with the particular choice of parameters
$\Gamma =\pm i\overline{M}/\sqrt{2}$, $\Upsilon=+M$.  Note that in terms
of the real charges this is $P=\pm\sqrt{2}m,Q=\pm\sqrt{2}l$, so that
this case reduces to an extremal magnetically charged black hole in the
$l=0$ limit.  With this choice for the charges eqn.
(\ref{stringmetric}) becomes, in terms of the radial coordinate
$\bar{r}$ defined above,
\begin{equation}
d\bar{s}^2 = |\bar{r}+2m+2il|^2\left\{
\frac{1}{(\bar{r}+2m)^2}(dt+2l\cos\theta d\varphi)^2
-\frac{1}{\bar{r}^2}d\bar{r}^2 -d\Omega^2\right\}\  ,
\end{equation}
which does have the form of an infinite throat as $\bar{r}\rightarrow
0$,
\begin{equation}
d\bar{s}^2 \approx  4(m^2+l^2)\left\{
\frac{1}{4m^2}( dt+2l\cos\theta d\varphi )^2 -
\frac{1}{\bar{r}^2}d\bar{r}^2 - d\Omega^2\right\} \, .
\end{equation}

There is a second interesting way to take the throat limit, analogous to
the `black hole plus throat' limit of the extremal magnetic black hole
\cite{GPS}.  To reach this limit, start again at (\ref{stringmetric})
and make the coordinate transformation $r=m+r_0(1+2\sinh ^2(\sigma/2))$.
Taking the extremal limit, $r_0\rightarrow 0$, with the charges fixed as
in the last paragraph, gives
\begin{equation}\label{johnsonthroat}
d\bar{s}^2 = 4(m^2+l^2)\left\{
\frac{\sinh ^2\sigma}{(2m\cosh\sigma +2\sqrt{m^2+l^2})^2}
( dt+2l\cos\theta d\varphi )^2  -d\sigma^2 -d\Omega^2\right\}.
\end{equation}
This limit of the extremal Taub-NUT solutions was recently found in an
exact conformal field theory construction by Johnson \cite{JOHNSON}.

The scalar curvature of the charged Taub-NUT solutions is given by
\begin{equation}
{\cal R} = \frac{2|\Upsilon|^2 (r-r+)(r-r_-)}{R^6}\ ,
\end{equation}
from which we see that the curvature is singular wherever $R^2$
vanishes.  From (\ref{taubnut}) we can see that this can happen, if
$|\Upsilon|^2\ge l^2$, which corresponds to $(Q^2+P^2)^2\ge
4l^2(m^2+l^2)$.  We see that there is a sort of competition between the
electromagnetic charge and the NUT charge in determining whether or not
the geometry is singular.  These singularities will occur at
$r_{sing}=\pm\sqrt{|\Upsilon|^2-l^2}$.  In the limit $l=0$, $r_{sing}$
coincides with $r_-$, giving the singular inner black hole horizon.  In
the extremal limit we have $r_+=r_-=r_{sing}=m$.  There may also in
certain cases be curvature singularities, such as the Schwarzschild one,
which do not contribute to the scalar curvature and which we have not
examined.


\section{IWP Solutions}

The IWP solutions in dilaton-axion gravity, like the IWP solutions in
Einstein-Maxwell theory, can be compactly expressed in terms of a single
complex function on the 3-dimensional plane.  In the present case, this
function gives the value of the complex dilaton-axion field $\lambda$,
which is independent of the time coordinate.  The metric and gauge field
are then given in terms of the real and imaginary parts of this function
by
\begin{equation}\label{iwpsolutions}
ds^{2}=e^{2\phi}(dt+\omega_{\hat{\imath}} dx^{\hat{\imath}})^{2}
-e^{-2\phi} d\vec{x}^{2},
\qquad
A_{\mu} =  \pm\frac{1}{\sqrt{2}}e^{2\phi}(1,\vec{\omega})\, ,
\end{equation}
where $d\vec{x}^{2}$ is the Euclidean 3-dimensional metric and the
components $\omega_{\hat{\imath}}$ of the 1-form
$\omega\equiv\omega_{\hat{\imath}} dx^{\hat{\imath}}=\vec{\omega}\cdot
d\vec{x}$ are solutions to the equation
\begin{equation}\label{curl}
\epsilon_{ijk}\partial_{\hat{\jmath}}\omega_{\hat{k}}=
-\partial_{\hat{\imath}} a\, .
\end{equation}
Given these relations between the fields, the
entire set of equations of motion reduce to the single equation
\begin{equation}\label{laplace}
\partial_{\hat{\imath}}\partial_{\hat{\imath}}\lambda =0\, ,
\end{equation}
i.e. the 3-d Euclidean Laplacian of $\lambda$ is zero.
Note that (\ref{laplace}) includes the integrability
condition for (\ref{curl}).  Given this, it is natural to define a
second $1$-form $\eta=\eta_{\hat{\imath}} dx^{\hat{\imath}}$ by
\begin{equation}\label{curl2}
\epsilon_{ijk}\partial_{\hat{\jmath}}\eta_{\hat{k}}=
-\partial_{\hat{\imath}} e^{-2\phi}\, .
\end{equation}
The $SL(2,R)$ dual gauge potential is then given by
\begin{equation}\label{dual}
\tilde{A}_\mu = \mp\frac{1}{\sqrt{2}}(ae^{2\phi},ae^{2\phi}\vec{\omega}
+\vec{\eta})\, .
\end{equation}
The scalar curvature of the IWP spacetimes is given, again as an
equation in the flat background metric, by
\begin{equation}
R = -2(\partial\phi)^2-{1\over 2} e^{4\phi}(\partial a)^2
  = -{1\over 2} \frac{\partial_{\hat{\imath}}\lambda
\partial_{\hat{\imath}}\overline{\lambda}}{({\mbox{Im}} \; \lambda)^{2}} \ .
\end{equation}
We see that there will be a scalar curvature singularity if ${\mbox{Im}}
\; \lambda =e^{-2\phi}= 0$ and $\partial_{\hat{\imath}}\lambda
\partial_{\hat{\imath}}\overline{\lambda}$ does not vanish sufficiently
rapidly.

There are many solutions to eqn.  (\ref{laplace}) corresponding to
different IWP solutions.  We will be primarily interested in solutions
which are asymptotically locally flat, and for which $\lambda$ is
singular only at isolated points in the $3$-dimensional plane.  We can
then write $\lambda(\vec{x})$ in the following form,
\begin{equation}
\lambda(\vec{x}) =i\left( 1 + \sum_{i=1}^N \frac{2(m_i+
in_i)}{r_i}\right )\, ,
\qquad
r_i^2=(x-x_i)^2 +(y-y_i)^2 +(z-z_i)^2\, ,
\end{equation}
where the parameters $m_i,n_i$ are real and correspond to the masses and
NUT charges of the objects.  The parameters $\vec{x}_i$ may be complex.

If we take all the $n_i=0$ and $\vec{x}_i$ real, then the axion $a$, and
hence also $\omega$, vanishes, and the IWP solution (\ref{iwpsolutions})
reduces to the static MP type solutions of ref. \cite{G} with pure
electric charge.  These correspond to collections of extremal charged
black holes.  As with the ordinary IWP solutions there are two basic
ways of moving away from the static limit; either by allowing the
positions $\vec{x}_i$ to have imaginary parts, or by taking some of the
$n_i\ne 0$.  The first option again introduces angular momentum and the
second NUT charge.  By exercising these two options in turn for a single
source, we will see that we recover special limits of known solutions.

For example, take $\lambda=i(1+2m/R)$ with $R^2=x^2+y^2+(z-i\alpha)^2$
$\alpha$ and $m$ are real parameters.  If we change to spheroidal
coordinates $(r,\theta,\varphi)$ given by
\begin{equation}\label{spheroidal}
x\pm iy = \sqrt{r^2 +\alpha^2} \sin\theta \exp(\pm i\varphi)\, ,
\qquad z=r\cos\theta .
\end{equation}
then $\lambda$ has the form
\begin{equation}
\lambda = i\left(1 + \frac{2M}{r +i\alpha \cos \theta}\right) =
\frac{2M\alpha\cos\theta}{r^2+\alpha^2\cos^2\theta} +i
\left(1+\frac{2Mr}{r^2 + \alpha^2\cos^2\theta}\right) \ ,
\end{equation}
and a solution of (\ref{curl}) is given by
\begin{equation}
\omega_\varphi = \frac{2 M\alpha r
\sin^2\theta}{r^2+\alpha^2\cos^2\theta} \ .
\end{equation}
The metric and gauge field are then determined to be
\begin{eqnarray}\label{senlimit}
ds^2 &=& \left( 1+\frac{2Mr}{r^2 + \alpha^2\cos^2\theta}\right)
\left( dt + \frac{2 M\alpha r \sin^2\theta}{r^2+\alpha^2\cos^2\theta}
d\varphi\right)^2 -\left( 1+\frac{2Mr}{r^2 +
\alpha^2\cos^2\theta}\right) d\vec{x}^{2}\, ,
\nonumber\\
d\vec{x}^{2} &=& \frac{r^2 +\alpha^2\cos^2\theta}{r^2+a^2}dr^2
+(r^2 +a^2\cos^2\theta)d\theta^2
+(r^2+a^2)\sin^2\theta d\varphi^2\, ,
\\
A_t &=& \pm\frac{r^2+\alpha^2\cos^2\theta}{\sqrt{2}(r^2 +\alpha^2
\cos^2\theta +2Mr)}\, ,
\nonumber\\
A_\varphi &=& \pm\frac{2M\alpha r\sin^2\theta}{r^2+\alpha^2\cos^2\theta
+2Mr}\  .
\nonumber
\end{eqnarray}
This coincides, after correcting for changes in convention and a shift
by $m$ in the radial coordinate, with Sen's rotating solution \cite{SEN}
in the limit in which the electric charge $q$ and mass $m$ are related
by $q=\sqrt{2}m$.  Hence, the situation is the same as with the
original IWP solutions.  The object has the same charge to mass ratio as
an extremal static black hole, but for $\alpha\ne 0$ corresponds to a
naked singularity.

Next consider the second option, taking $\lambda = i [ 1 +2(m+il)/r]$,
where $m,l$ are real and $r$ is the usual spherical radial coordinate.
The metric and gauge field are then given by
\begin{eqnarray}
ds^2 &=& (1+\frac{2m}{r})^{-1}
(dt +2l\cos\theta d\varphi)^2 - (1+\frac{2m}{r})(dr^2 + r^2
d\Omega^2)\, ,
\nonumber\\
A_t &=& \pm\frac{1}{\sqrt{2}}(1-\frac{2m}{r})\, , \\
A_\varphi &=& \pm\frac{1}{\sqrt{2}}(1-\frac{2m}{r})2l\cos\theta\, .
\nonumber
\end{eqnarray}
This corresponds to the extremal charged Taub-NUT solution given in
the last section, with a particular choice made for the charges.

Finally, it is clear that the IWP solutions presented in this section are not
the most general ones.  They do not have the property of manifest $SL(2,R)$
duality.
As stated above, the present solutions reduce, in the static limit, to the
MP solutions with pure electric charge.  Given the $SL(2,R)$ invariance of the
equations of motion, one can certainly find a more general class of IWP
solutions, which would reduce in the static limit to the generally charged
multi-black hole solution given in \cite{KOSTATIC}.  Amongst these more general
solutions, there will be examples, including the extremal, magnetic Taub-NUT
solution already given above,
for which the corresponding string metrics will
have infinite throats.


\section{Supersymmetry of dilaton-axion IWP Solutions}

So far, in dilaton-axion gravity, the only solutions known to have
unbroken supersymmetries were the extreme static dilaton-axion
black holes \cite{US,TOdual,KOSTATIC,TOMAS}. It is natural to ask now
if the dilaton-axion IWP metrics we have presented in the previous
section are supersymmetric. The analogy with Einstein-Maxwell IWP
solutions \cite{T} would suggest that the answer to this question is
affirmative.

Before proceeding to check this statement let us explain its meaning.
The action eqn.  (\ref{action}) can be considered as a consistent
truncation of $N=4$, $d=4$ ungauged supergravity where all the fermionic
fields (and some bosonic fields as well) are set to zero.  The solutions we
have presented can therefore be considered as solutions of $N=4$, $d=4$
ungauged supergravity with all the fermionic fields set to zero. it is
reasonable to ask now whether these solutions are invariant under local
supersymmetry transformations.  Since the bosonic supersymmetry
transformation rules
are proportional to the (vanishing) fermionic fields, the bosonic fields
are always invariant.  However, the possibility remains that after a
local supersymmetry transformation non-vanishing fermion fields appear.
In fact this is what generally happens.  For very special configurations
there are a finite number of local supersymmetry transformations that
leave the fermionic part of the solution vanishing.  In this case one
says that the configurations are supersymmetric (i.e. have unbroken
supersymmetries) and the corresponding supersymmetry parameters are
called Killing spinors.

The $N=4$, $d=4$ fermionic supersymmetry rules are
\begin{eqnarray}
{\textstyle\frac{1}{2}}\delta_{\epsilon}\Psi_{\mu I} & = &
\nabla_{\mu}\epsilon_{I}
-{\textstyle\frac{i}{4}}e^{2\phi}\partial_{\mu}a\epsilon_{I}
-{\textstyle\frac{1}{2\sqrt{2}}}e^{-\phi}\sigma^{ab}
F_{ab}^{+}\gamma_{\mu}\alpha_{IJ}\epsilon^{J}\; ,
\label{eq:susytrans1}
\\
\nonumber \\
{\textstyle\frac{1}{2}}\delta_{\epsilon}\Lambda_{I} & = &
-{\textstyle\frac{i}{2}}(e^{2\phi}\!\not\!\partial\lambda)\epsilon_{I}
+{\textstyle\frac{1}{\sqrt{2}}}e^{-\phi}\sigma^{ab}
F_{ab}^{-}\alpha_{IJ}\epsilon^{J}\; .
\label{eq:susytrans2}
\end{eqnarray}
The statement that the IWP solutions are supersymmetric then means that
the equations $\delta_{\epsilon}\Psi_{\mu I}=0\,
,\,\, \delta_{\epsilon}\Lambda_{I}=0$ have
at least one set of solutions $\epsilon_{I}\neq 0$ when we substitute the
IWP fields into eqns.  (\ref{eq:susytrans1}), (\ref{eq:susytrans2}).

To show that this is indeed the case,
we first calculate the self-dual spin connections and
self-dual vector field strengths.  A basis of vierbeins for
the metric (\ref{iwpsolutions}) is provided by the $1$-forms and
vectors
\begin{eqnarray}
e^{0} & = & e^{\phi}(dt+\omega)\, ,
\nonumber \\
e^{i} & = & e^{-\phi} dx^i\, ,
\nonumber \\
e_{0} & = & e^{-\phi}\partial_{\hat{0}}\, ,
\nonumber \\
e_{i} & = &
e^{\phi}[-\omega_{i}\partial_{\hat{0}}+\partial_{\hat{\imath}}]\, .
\end{eqnarray}
The self-dual (in the upper Lorentz indices) part of the spin
connection $1$-form is given by
\begin{eqnarray}
\omega^{+0i} & = &
{\textstyle\frac{i}{4}} e^{3\phi}
[\partial_{\hat{\imath}}\overline{\lambda}e^{0}
+i\epsilon_{ijk}\partial_{\hat{\jmath}}\overline{\lambda} e^{k}]\, ,
\nonumber \\
\nonumber \\
\omega^{+ij} & = &
{\textstyle\frac{1}{4}} e^{3\phi}
[\epsilon_{ijk}\partial_{\hat{k}}\overline{\lambda } e^{0}+
2i\partial_{[\hat{\imath}}\overline{\lambda}\delta_{j]k}e^{k}]\, .
\label{spinconn}
\end{eqnarray}
The different components of the self-dual part of the electromagnetic
tensor $F$ are
\begin{eqnarray}
F^{+}_{0i} & = &
\pm{\textstyle\frac{i}{2\sqrt{2}}} e^{4\phi}
\partial_{\hat{\imath}}\overline{\lambda}\, ,
\nonumber \\
\nonumber \\
F^{+}_{ij} & = &
\mp{\textstyle\frac{1}{2\sqrt{2}}} e^{4\phi} \epsilon_{ijk}
\partial_{\hat{k}}\overline{\lambda}\, .
\label{maxw}
\end{eqnarray}
Now, let us examine the dilatino supersymmetry rule eqn.
(\ref{eq:susytrans2}). First observe that
\begin{eqnarray}
-{\textstyle\frac{i}{2}} e^{2\phi} \not\!\partial\lambda & = &
\gamma^{i}(-{\textstyle\frac{i}{2}} e^{3\phi}
\partial_{\hat{\imath}}\lambda)\, ,
\nonumber \\
{\textstyle\frac{1}{\sqrt{2}}} e^{-\phi} \sigma^{ab}F_{ab}^{-}
\alpha_{IJ}\epsilon^{J}
& = & \gamma^{i}(-{\textstyle\sqrt{2}}F_{0i}^{-}\alpha_{IJ}
\gamma^{0}\epsilon^{J})=
\nonumber \\
& = &
\gamma^{i}(\pm{\textstyle\frac{i}{2}} e^{3\phi}
\partial_{\hat{\imath}}\lambda \alpha_{IJ}\gamma^{0}\epsilon^{J})\, .
\end{eqnarray}
All this implies
\begin{equation}
{\textstyle\frac{1}{2}}\delta_{\epsilon}\Lambda_{I}=
-{\textstyle\frac{i}{2}} e^{2\phi} \!\not\!\partial\lambda
[\epsilon_{I}\mp\alpha_{IJ}\gamma^{0}\epsilon^{J}]= 0\, ,
\end{equation}
which gives the constraint on the Killing spinors
\begin{equation}\label{eq:constraint}
\epsilon_{I}\mp\alpha_{IJ}\gamma^{0}\epsilon^{J}= 0\, .
\end{equation}
This is a first  clear indication that supersymmetry may be working.
In the dilatino part the spin connection does not enter, it was the
relation between the dilaton and vector field which was relevant.

Now we take the $\hat{0}$-component of the gravitino supersymmetry rule
Eq.  (\ref{eq:susytrans1}) imposing time-independence of the Killing
spinor set $\epsilon_{I}$, i.e.
\begin{equation}
\partial_{\hat{0}}\epsilon_{I}=0\, .
\end{equation}
The equation is
\begin{equation}
-{\textstyle\frac{1}{2}}\omega^{+ab}_{\hat{0}}\sigma_{ab}\epsilon_{I}
-{\textstyle\frac{1}{2\sqrt{2}}}e^{-\phi}\sigma^{ab}F^{+}_{ab}
\alpha_{IJ}\gamma_{\hat{0}}\epsilon_{J}= 0\, .
\end{equation}
First, using equation (\ref{spinconn}), we get
\begin{equation}
-{\textstyle\frac{1}{2}} \omega^{+ab}_{\hat{0}} \sigma_{ab} \epsilon_{I}
= \gamma^{i}(-{\textstyle\frac{i}{4}} e^{3\phi}
\partial_{\hat{\imath}}\overline{\lambda}\gamma^{0}\epsilon_{I})\, .
\end{equation}
Using eq. (\ref{maxw}), we have
\begin{equation}
-{\textstyle\frac{1}{2\sqrt{2}}}e^{-\phi}\sigma^{ab}F^{+}_{ab}
\alpha_{IJ}\gamma_{\hat{0}}\epsilon_{J}
= \gamma^{i}(\mp{\textstyle\frac{i}{4}} e^{3\phi}
\partial_{\hat{\imath}}\overline{\lambda}\alpha_{IJ}\epsilon_{J})\, .
\end{equation}
Putting both pieces together we get
\begin{equation}
{\textstyle\frac{1}{2}} \delta_{\epsilon} \Psi_{\hat{0} I}=
-{\textstyle\frac{i}{4}} e^{2\phi} \!\not\!\partial \overline{\lambda}
\gamma^{0} [\epsilon_{I}\mp\alpha_{IJ}\gamma^{0}\epsilon^{J}]=0\,  ,
\end{equation}
which is obviously satisfied on account of the constraint given in
eq. (\ref{eq:constraint}).

The last part is to verify the space component of the gravitino
supersymmetry transformation.  This is a differential equation on the
Killing spinor $\epsilon_{I}$ which has the form
\begin{equation}
{\textstyle\frac{1}{2}} \delta_{\epsilon} \Psi_{\hat{\imath} I}=
\partial_{\hat{\imath}} \epsilon_{I}
-{\textstyle\frac{1}{2}} \omega_{\hat{\imath}}^{+ab} \sigma_{ab}
\epsilon_{I}
-{\textstyle\frac{i}{4}} e^{2\phi} \partial_{\hat{\imath}}a \epsilon_{I}
-{\textstyle\frac{1}{2\sqrt{2}}} e^{-\phi} \sigma^{ab}
F_{ab}^{+} \gamma_{\hat{\imath}} \alpha_{IJ} \epsilon^{J}=0\  .
\label{eq:that}
\end{equation}
First we have
\begin{equation}
-{\textstyle\frac{1}{2}} \omega_{\hat{\imath}}^{+}{}^{ab}
\sigma_{ab} \epsilon_{I}
=\gamma^{j} [-e^{\phi} \omega_{\hat{\imath}} \omega_{0}^{+}{}^{0j}
-e^{-\phi} \omega_{i}^{+}{}^{0j}] \gamma^{0} \epsilon_{I}\, ,
\end{equation}
and secondly
\begin{equation}
-{\textstyle\frac{1}{2\sqrt{2}}} e^{-\phi} \sigma^{ab}
F^{+}_{ab} \alpha_{IJ} \gamma_{\hat{\imath}} \epsilon_{J}
=\gamma^{j} [-{\textstyle\frac{1}{\sqrt{2}}}
\omega_{\hat{\imath}} F^{+}_{0j}
+{\textstyle\frac{1}{\sqrt{2}}} e^{-2\phi} F^{+}_{0j} \gamma^{i}
\gamma^{0}] \alpha_{IJ}\epsilon_{J}\, .
\end{equation}
Now we sum the last two equations and use our last result (the $\hat{0}$
component of the gravitino supersymmetry transformation rule) and the
constraint Eq.  (\ref{eq:constraint}) to simplify the sum.  We get
\begin{equation}
-e^{-\phi} \gamma^{j} [\omega_{i}^{+}{}^{0j} \gamma^{0}-
{\textstyle\frac{1}{\sqrt{2}}}
e^{-\phi} F_{0j}^{+} \gamma^{i}] \epsilon_{I}\, .
\label{eq:this}
\end{equation}
To further simplify this equation we use the identity
\begin{equation}
\gamma^{j} \gamma^{i} F_{0j}^{+}=
F_{0i}^{+} +\gamma^{j} F_{ij}^{+} \gamma^{0} \gamma_{5}\, .
\end{equation}
Substituting it into Eq. (\ref{eq:this}), noting that
$\epsilon_{I}$ has negative chirality and using the explicit form of the
spin connection  and the electromagnetic field of the IWP solutions
eqns. (\ref{spinconn}) and (\ref{maxw}) we get
\begin{equation}
-{\textstyle\frac{1}{2}} \omega_{\hat{\imath}}^{+ab}
\sigma_{ab} \epsilon_{I}
-{\textstyle\frac{1}{2\sqrt{2}}}e^{-\phi}\sigma^{ab}
F_{ab}^{+} \gamma_{\hat{\imath}} \alpha_{IJ} \epsilon^{J}=
\pm{\textstyle\frac{i}{4}} e^{2\phi} \partial_{\hat{\imath}}
\overline{\lambda}\epsilon_{I}\, .
\end{equation}
This can be substituted back into Eq. (\ref{eq:that}) getting, at
last, a much simpler differential equation for the Killing spinors
\begin{equation}
\partial_{\hat{\imath}}\epsilon_{I}+{\textstyle\frac{1}{4}}
e^{2\phi}(\partial_{\hat{\imath}}e^{e^{-2\phi}})\epsilon_{I}=0\, ,
\end{equation}
which can be rewritten as
\begin{equation}
\partial_{\hat{\imath}} (e^{e^{-\phi/2}}\epsilon_{I})=0\  .
\end{equation}
Thus, the Killing spinor set $\epsilon_{I}$ exists and is given in
terms of a set of constant spinors $\epsilon_{I\, (0)}$ by
\begin{equation}
\epsilon_{I}=e^{\phi/2}\epsilon_{I\, (0)}\  ,
\end{equation}
where the constant spinors satisfy the same constraint
Eq. (\ref{eq:constraint}) as the Killing spinors themselves:
\begin{equation}
\epsilon_{I\, (0)}\mp\alpha_{IJ}\gamma^{0}\epsilon^{J}_{(0)}=0\  .
\end{equation}
This constraint limits the number of independent components of the
Killing spinor set $\epsilon_{I}$ to half of the total.  Thus, the
dilaton-axion IWP solutions have two unbroken supersymmetries when
embedded into $N=4$, $d=4$ ungauged supergravity.

Let us now analyze this result.  It is  known that supersymmetry
enforces a Bogomolnyi-Gibbons-Hull (BGH) bound  \cite{GH} on the charges,
which is saturated by those configurations having unbroken
supersymmetries.  These charges are defined at asymptotic infinity
and usually consist of the mass and central charges of the supersymmetry
algebra (in the case of extended supersymmetry theories).
For static, asymptotically flat black holes,
there is a fascinating coincidence
between the BGH bound, which implies supersymmetry,
and the extremal bound, which implies the
absence of naked singularities \cite{US}.
If one tries to extend this
to configurations which are not asymptotically flat or are not static, one
finds that the rule no longer holds.  For instance, static asymptotically
anti-deSitter solutions with both unbroken supersymmetries are known to have
naked singularities \cite{R}.  Likewise, there are Einstein-Maxwell IWP
metrics which are both supersymmetric and have naked singularities.
In this latter case (take for instance the Kerr-Newman
metric with $Q^{2}=M^{2}$) the difficulty can be traced to the fact that
the angular momentum of the hole is present
in the condition for the absence of naked
singularities, but not in the BGH bound.

For the IWP solutions, NUT charge plays an interesting role in the analysis
of BGH bounds as
well.  We discuss this briefly here for our dilaton-axion solutions, but a
parallel discussion would hold for the Einstein-Maxwell IWP solutions and,
to our knowledge, has not previously been given.  We hope to return to a more
complete discussion of this point in future work.
The IWP solutions being supersymmetric, must saturate some BGH
bound.  One might naively think that, the supersymmetry algebra being the same,
the BGH bound should be the same as that derived in \cite{TOdual}
for the extreme static dilaton-axion black holes, namely
\begin{equation}
m^{2}+|\Upsilon|^{2}-4|\Gamma|^{2}\geq 0\, .
\end{equation}
However, from the discussion in the previous sections, it is clear that
the correct BGH bound saturated by the IWP solutions is
\begin{equation}
m^{2}+l^{2}+|\Upsilon|^{2}-4|\Gamma|^{2}\geq 0\, .
\label{eq:BGH}
\end{equation}
The NUT charge $l$ appears here as a sort of dual to the usual ADM mass.
At first, this is a bit surprising.  However, the NUT charge is well known to
play such a role; Taub-NUT space being interpreted as a gravitational dyon
(see e.g. \cite{DOWKER} and references therein).
If asymptotic conditions are relaxed to allow for nonzero NUT charge, then
the NUT charge must arise as a boundary term in a positive energy construction
on a spatial slice.

Finally, note that once again the angular momentum does not appear in the bound
(\ref{eq:BGH}) and that, in the cases in which the NUT charge vanishes
and we have asymptotically flat geometries in the usual sense,
we will have the same problems discussed above.

{}From the dilaton-axion IWP solutions, which we have presented in
the previous section, one can generate more general solutions using
$SL(2,R)$-duality rotations \cite{STW,S}. The new solutions will have
the same Einstein-frame metric but different dilaton, axion and vector
fields, and, therefore, different string-frame metric. In addition they
will have the same number of unbroken supersymmetries
\cite{TOdual,TOMAS}. The same problems with supersymmetry and naked
singularities will be present in the $SL(2,R)$-rotated solutions.

The relation between extreme  black holes of
ordinary
Einstein-Maxwell theory and IWP metrics, which are both
supersymmetric at
the classical level when embedded into $N=2$ supergravity, was
analysed in
\cite{T}, \cite{TRACE}. In particular the Killing spinors of the IWP
solution  depend on space coordinates through some complex harmonic function
$V$.\begin{equation}
\epsilon_{I}(x) =V^{1/2} (x)\epsilon_{I\ (0)}\, , \qquad I=1,2.
\label{V}\end{equation}
where $\epsilon_{I\ (0)}$  are some constant spinors.  The same  function
$V$  is also used in the ansatz for the metric and for the vector field.
If we choose the imaginary part of the function $V$ to become zero, we reduce
the
IWP solution to the extreme electrically charged Reissner-Nordstr\"om solution.
Simultaneously the Killing spinor dependence on the space-time is reduced to
the
dependence on the real function $V$ in eq. (\ref{V}).

In $N=4$ supergravity the Killing spinors of $SL(2, R)$-form invariant
dilaton-axion black holes also
have
a dependence on space coordinates through a complex function
\cite{TOdual,TOMAS}. The real part
of this function is defined by the $g_{tt}$ component of the metric and the
imaginary part of it is related to the $Arg (\gamma \lambda + \delta)$,
where $\lambda$ is the dilaton-axion field and $ \gamma , \delta $ are
parameters of $SL(2, R)$ transformations defined in equation (\ref{SL}).

We have found that the Killing spinors for dilaton-axion
IWP solutions, described in the previous sections of this paper, depend on
space coordinates through a real function $e^{{1\over 2} \phi}$. Such
dependence
is known to exist for pure electric dilaton black holes \cite{US}.  If we would
consider the
more general class of dilaton-axion
IWP solutions, related to those presented above by a
generic $SL(2, R)$ transformations, we would get a complex function, defining
the dependence of a Killing spinor on space coordinates \cite{TOdual,TOMAS}.
However here we have studied the special form of IWP solutions, whose
Killing spinors depend on $x$ only through a dilaton field.
  In stringy frame electric black holes as well as our IWP solutions have
Killing spinors with
 dependence on space coordinates in the form $e^{ \phi}$. This gives us a nice
bridge to the
supersymmetric
gravitational waves and their dual partners. We have studied such dependence
before
and know that the relevant gravitational waves have constant Killing spinors
\cite{BKO}.
The Killing
spinors of the dual waves have the following dependence on space coordinates
 \cite{BEK}
\begin{equation}
\epsilon_{str}(x) = e^{ \phi (x)} \epsilon_{0}\ .
\end{equation}


\section{IWP Solutions From Stringy Waves}

The previous section shows that to find the unbroken  supersymmetry of the
4-dimensional IWP dilaton-axion solutions one has  to solve a rather involved
set of equations. In the present situation,
where the result poses a conceptual puzzle, it is useful to have
an independent means of checking supersymmetry.

Interesting enough,  there does exist a completely independent method of
deriving the general IWP dilaton-axion solutions, which automatically
insures that they are supersymmetric.  There exist gravitational wave
solutions in 10-dimensional $N=1$ supergravity, which have been shown to
be supersymmetric \cite{BKO} and which are known as supersymmetric
string waves (SSW's).  Further, it was recently shown \cite{BKO2} that
certain SSW's are dual, under a sigma model duality transformation \cite{BUS},
to
extremal 4-dimensional black holes, embedded in 10-dimensional geometry.
Here, we show that certain other SSW's transform under sigma model
duality into 4-dimensional IWP solutions embedded into 10-dimensional
geometry.  One can preserve the unbroken supersymmetry of higher-dimensional
solutions by using dimensional reduction of supergravity \cite{Cham}. The fact
that sigma-model duality preserves supersymmetry is proved in \cite{BKO3}. Also
 a direct proof of supersymmetry of dual partners of the wave solution in
10-dimensional theory
is available \cite{BEK}.

Here we recount
the construction briefly.  More details can be found in \cite{BKO2}. Our
conventions in this section are those in  \cite{BEK},  \cite{BKO2} .

We consider the zero slope limit of the effective string action.  This
limit corresponds to 10-dimensional $N=1$ supergravity.  The Yang-Mills
multiplet will appear in first order $\alpha'$ string corrections.  The
bosonic part of the action is
\begin{equation}
S=\frac{1}{2}\int d^{10}x
e^{-2 \phi}\sqrt{- g} \, [ -{\cal R}+
(\partial\phi )^{2}-\frac{3}{4}H^{2}]\, ,
\label{eq:actionD1}
\end{equation}
where the 10-dimensional fields are the metric $g_{MN}$, the
3-form field strength $H_{MNL}=\partial_{[M} B_{NL]}$ and the dilaton $\phi$.
The zero slope
limit of the SSW solutions \cite{BKO} in $d=10$ are given by the
Brinkmann metric and 2-form gauge potential
\begin{eqnarray}
\label{SSW}
ds^2 &=& 2 d \tilde u d \tilde v + 2A_M d\tilde x^M d\tilde u -
 \sum_{i=1}^{i=8} d\tilde x^id \tilde x^i\ ,\nonumber \\
\label{eq:SSW2}
B &=& 2 A_M d\tilde x^M \wedge d\tilde u\ , \qquad
A_{v} =0 \, ,
\end{eqnarray}
where the indices run over the values $i=1,\dots , 8, \; M = 0, 1, \dots
, 8, 9$ and we are using the notation
$x^M=\{\tilde{u},\tilde{v},\tilde{x}^i\}$ for the 10-dimensional
coordinates.  We have put the tilde over the 10-dimensional coordinates
for this solution, because we will have to compare this original
10-dimensional solution after dual rotation with the 4-dimensional one,
embedded into the 10-dimensional space.  A rather non-trivial
identification of coordinates, describing these solutions will be
required.

The equations of motion reduce to equations for
$A_u(\tilde x^i)$ and $A_i(\tilde x^j)$, which are
\begin{equation}
\label{eq:Lapl}
\triangle A_u = 0\ , \hskip 1.5truecm \triangle\partial^{[i}A^{j]} =
0\ ,
\end{equation}
where the Laplacian $\triangle$ is taken over the transverse directions only.

Application of a sigma-model duality transformation \cite{BUS} to the
SSW solutions given in eq.~(\ref{SSW}) leads to the following new
supersymmetric solutions of the equations of motion in the zero slope
limit \cite{BEK}:
\begin{eqnarray}
\label{dualwave}
ds^2 & = &  2e^{2\phi}\bigl \{ d\tilde ud\tilde v +  A_i d\tilde
ud\tilde x^i  \bigr \} - \sum_{i=1}^{i=8} d\tilde x^id\tilde x^i\, ,
\nonumber\\
B  & = & -2 e^{2\phi} \bigl\{  A_u d\tilde u \wedge d\tilde v +
A_id\tilde u \wedge d\tilde x^i \bigr \}\, ,
\\
e^{-2\phi}  &=& 1 - A_u \, ,
\nonumber
\end{eqnarray}
where, as before, the functions $A_{M} = \{A_u= A_u (\tilde x^j), A_v=0,
A_i = A_i (\tilde x^j)\}$ satisfy eqns.  (\ref{eq:Lapl}).  We call this
new solution the dual partner of the SSW, or `dual wave' for simplicity.

The next step after the dual rotation of the wave is to dimensionally
reduce from $d=10$ to $d=4$.  Both steps can be performed in a way which
keeps the unbroken supersymmetry of the original SSW configuration
intact.  The rules for this procedure have been worked out in refs.
\cite{BKO2} and \cite{BKO3}, and more details of the procedure
may be found there.

The 4-dimensional action, related to (\ref{eq:actionD1}) by our
dimensional reduction is\footnote{Note that the 4-vector field $V_{\mu}$
in this action is related to the 4-dimensional 4-vector field $A_{\mu}$
in the action (\ref{action}) as $V_{\mu}= \sqrt 2 A_{\mu}$. The difference in
notation
between the 4-dimensional action  (\ref{action}) and equation
below is explained  in \cite{BKO3}.}
\begin{equation}\label{eq:action4trunc}
S=\frac{1}{2}\int d^{4}x
e^{-2\phi}\sqrt{-g}[-R+4(\partial\phi)^{2}-\frac{3}{4}H^{2}
+\frac{1}{2}F^{2}(V)]\, ,
\end{equation}
where
\begin{eqnarray}\label{3form}
F_{\mu\nu}(V) & = & 2\partial_{[\mu}V_{\nu]}\, ,
\nonumber \\
H_{\mu\nu\rho} & = & \partial_{[\mu}B_{\nu\rho]}+
V_{[\mu}F_{\nu\rho]}(V)\, .
\end{eqnarray}
The embedding of the $4$-dimensional fields in this action in $d=10$
are as follows:
\begin{eqnarray}\label{eq:uplift}
g^{(10)}_{\mu\nu} & = & g_{\mu\nu}-V_{\mu}V_{\nu}\, ,
\nonumber \\
g^{(10)}_{4\nu} & = & - V_{\nu}\, ,
\nonumber \\
g^{(10)}_{44} & = & -1\, ,
\nonumber \\
g^{(10)}_{IJ} & = & \eta_{IJ}=-\delta_{IJ}\, ,
\nonumber \\
B^{(10)}_{\mu\nu} & = & B_{\mu\nu}\, ,
\nonumber \\
B^{(10)}_{4\nu} & = & V_{\nu}\, ,
\nonumber \\
\phi^{(10)} & = & \phi\, ,
\end{eqnarray}
where the
indices $\mu,\nu$ run over the range $0,1,2,3$ and $I,J$ run over the
range $5,6,7,8,9$.  This formula can be used to uplift a $U(1)$
$4$-dimensional field configuration, including the dilaton and axion
fields, to a $10$-dimensional field configuration, in a way which is
consistent with supersymmetry.  Or vice versa, one may bring down to $4$
dimensions a 10-dimensional configuration which
has the property $ g^{(10)}_{4 \nu }= - B^{(10)}_{4 \nu }$.

To generate a dual-wave, which upon dimensional reduction gives the
$4$-dimensional IWP solutions, we make the following choices for the
vector $A_M$ in the $10$-dimensional SSW (\ref{SSW}).  The components
$A_M$ are taken to depend only on three of the transverse coordinates,
$\tilde{x}^1, \tilde{x}^2, \tilde{x}^3$, which will ultimately correspond to
our 3-dimensional space.  We choose one component, {\it e.g.}, $A_4$ to be
related to according to $A_4= \xi A_u$, where $\xi^2=\pm1$ depending on
the signature of spacetime\footnote{See \cite{BKO2} for the details.}.  Recall
that from (\ref{dualwave}) $A_u$ is
related to the dilaton by $A_u = 1-e^{-2\phi} $. For the remaining
components, we take only $A_1,A_2,A_3$ to be non-vanishing, and we
relabel these as $\omega_1,\omega_2,\omega_3$ for obvious reasons.  To
summarize, we then have
\begin{equation}
  A_4 = \xi A_u \ , \qquad A_1 \equiv \omega_1\ , \qquad A_2 \equiv
\omega_2\ , \qquad A_3 \equiv \omega_3\ ,
 \qquad A_5 = \dots = A_8=0.
\end{equation}
The equations of motion (\ref{eq:actionD1}), in terms of these degrees of
freedom, become
\begin{equation}\label{laplace3}
\triangle e^{-2\phi} = 0\ , \hskip 1.5truecm
\triangle\partial^{[i}\omega^{j]}=0\ ,
\end{equation}
where again the Laplacian is taken over the transverse directions only.

A few more steps are required to dimensionally reduce the dual wave and
recover the IWP solutions of the previous sections.  These include a
coordinate change,
\begin{equation}
\hat{x}=\tilde{x}^{4} + \xi \tilde{u}\,  , \qquad
\hat  v = \tilde v  + \xi  \tilde x^4 \ ,
\end{equation}
shifting $B$ by a constant value and a particular identification of the
coordinates in the dual wave solution with those in the uplifted IWP
solution,
\begin{eqnarray}\label{dualwave2}
t & = & \hat{v} =  \tilde v +\xi  \tilde x^4\, ,
\nonumber \\
x^{4} & = & \tilde{u}\, ,
\nonumber \\
x^{9} & = & \hat{x}= \tilde{x}^{4} + \xi \tilde{u}\, ,
\nonumber \\
x^{1,2,3,5,\dots,8} & =  & \tilde{x}^{1,2,3,5,\dots,8}\, ,
\end{eqnarray}
After all these steps our 10-dimensional dual wave becomes
\begin{eqnarray}
ds^{2} & = & 2e^{2\phi} d x^4 (dt   +\vec \omega \cdot d \vec  x) -
\sum_4^9 d x^{i}d x^{i} -d \vec x^2\, ,
\nonumber \\
B & = & - 2e^{2\phi} dx^4  \wedge  (dt +\vec \omega \cdot d \vec x) \, .
\end{eqnarray}
To recognize this as the lifted IWP solution, add and subtract from the
metric the term $e^{4\phi} (dt +\vec \omega \cdot d \vec x)^2$.  We can
then rewrite the dual wave metric (\ref{dualwave2}) as
\begin{equation}\label{eq:IPWwave}
ds^2 =  e^{4\phi} (dt +\vec \omega \cdot d \vec  x) ^2 -
 d \vec x^2 - \left(dx^4  -e^{2\phi} (dt + \vec \omega \cdot d
\vec x)\right)^2  -  \sum_5^9 d x^{i}d x^{i} \ .
\end{equation}
The first two terms now give the string metric for the $4$-dimensional
IWP solutions.  The non-diagonal components $g_{\mu 4}$, in the third
term, are interpreted as the $4$-dimensional gauge field components,
showing the Kaluza-Klein origin of the gauge field in this construction.
Note that the $4$-dimensional vector field components are also equal to
the off-diagonal components of the 2-form gauge field, giving the
overall identifications
\begin{equation}
g_{t 4 } = B_{t 4} =  e^{2\phi}= -V_t \ , \qquad g_{i 4} = B_{i 4}=
e^{2\phi} \omega_i= -V_i \ .
\end{equation}
The dilaton of the IWP solution, is identified with the fundamental
dilaton of string theory, rather than with one of the modulus fields.
The axion is identified with the $4$-dimensional part of the 3-form
field strength $H$ given in eq.  (\ref{3form}).  Note that these
components of $H$ come totally from the second term in (\ref{3form}),
since the 4-dimensional $B_{\mu\nu}$ vanishes.  The equations of motion
(\ref{laplace3}) coincide with the equation (\ref{laplace}) which we had
earlier for the IWP solutions.

This provides an independent proof of unbroken supersymmetry of
dilaton-axion IWP solutions.  Here we have demonstrated only the
relation between the SSW's and IWP solutions.  The proof that the
unbroken supersymmetry survives the duality transformation and special
compactification is the subject of other publications \cite{BKO2},
\cite{BKO3}.


\section{Discussion}

We have presented new, stationary, black hole type solutions to
dilaton-axion gravity.  These new solutions share many basic properties
with their counterparts in Einstein-Maxwell theory, while differing from
these counterparts in ways which are familiar from previous studies of
dilaton-axion black holes.  For the Taub-NUT solutions, two particularly
interesting features are the singularities which arise for charge
sufficiently large and the existence of an infinite throat for the
string metric in the extremal limit with predominantly magnetic charge.
This latter feature makes contact with recently derived exact CFT
results \cite{JOHNSON}.  It should be possible to find a more general
stationary solution in dilaton-axion gravity, analogous to the general
type D metric of Einstein-Maxwell theory \cite{DP}, including mass, NUT
charge, electromagnetic charge, rotation and an acceleration parameter.
This grand solution would encompass the present Taub-NUT solutions, as
well as the rotating solutions \cite{SEN} and C-metric type solutions
\cite{CMETRIC}, as subclasses.

As explained in the introduction, the discovery of the new,
supersymmetric, dilaton-axion IWP solutions brings back an apparent
paradox, to which a resolution had recently been posed in the case of
Einstein-Maxwell theory.  The paradox arises if one interprets unbroken
supersymmetry of a bosonic solution as indicating that the solution is a
physical ground state of the theory.  In Einstein-Maxwell theory the
fact that the IWP solutions are supersymmetric implies a large
degeneracy of the ground state in a given charge sector; single object
solutions with vanishing NUT parameter, for example, satisfy
$Q^2+P^2=M^2$ with arbitrary angular momenta.  This degeneracy, however,
is removed when quantum corrections are taken into account.  The trace
anomaly of $N=2$ supergravity is found to contradict the integrability
condition for supersymmetry for nonzero angular momentum \cite{TRACE}.
This provides a simple possible resolution of the paradox, since the
trace anomaly, related to the Gauss-Bonnet Lagrangian, is well known in
supergravity, being gauge independent and finite.  Any other quantum
corrections in gravitational theories are significantly more difficult
to control.  The gauge independent trace anomaly, however, vanishes in
the present case of $N=4$ supergravity, making it much more difficult to
understand any effect of quantum corrections in removing the degeneracy
of the supersymmetric state.  This will be the subject of future
investigations.

The fact that the new stationary solutions of $N=4$ supersymmetric
theory have naked singularities in the canonical geometry does not
contradict the conjecture \cite{US} that supersymmetry may act as a
cosmic censor for static, asymptotically flat configurations.  In $N=2$
theory, it was possible to establish that the more general stationary
solutions are not actually supersymmetric, when quantum corrections are
taken into account.  At present, we do not know whether any analogous
situation may hold in $N=4$ theory.

The presence of a fundamental dilaton field in our theory may be
considered as the source of a completely different interpretation of the
geometry.  In Einstein-Maxwell theory there is no natural source of a
conformal transformation to another metric.  However, in string theory
the natural geometry of the target space is the so-called stringy frame,
where the string metric is related to the canonical one by the Weyl
transformation $e^{2\phi}$.  Moreover, in string theory the
4-dimensional spacetime is not the only relevant space for understanding
of configurations which solve the equations of motion of the full
ten-dimensional theory.  A possible radical point of view in seeking a
resolution to our paradox would be to ignore the properties of the
canonical four-dimensional configuration and to study our solutions
instead in the stringy frame by ``lifting'' them up to ten dimensions.
We have done this in section 6 and shown that the IWP solutions are dual
partners of the supersymmetric string waves, which were found in
\cite{BKO}.  We hope to return in future work to the implications of
this dual relationship between black-hole-type solutions and gravitational
waves.

\section*{Acknowledgements}

We are grateful to Fay Dowker, Jerome Gauntlett,
Gary Gibbons, Paul Tod and Jennie Traschen for stimulating discussions.
We have heard from D. Galtsov that he has also found
some of the stationary solutions of axion-dilaton gravity presented here.
The work of R. K. was supported by NSF grant PHY-8612280 and the work of
T. O. was supported by an European Union Human Capital and
Mobility programme grant.


\end{document}